\documentclass[ALICE,manyauthors]{cernphprep}
\usepackage[comma,square,numbers,sort&compress]{natbib}
\usepackage{lineno}
\usepackage{xspace}
\usepackage{xcolor}
\usepackage{euscript}
\usepackage{graphicx}% Include figure files
\usepackage{dcolumn}% Align table columns on decimal point
\usepackage{bm}% bold math
\usepackage{graphics}
\usepackage{hyperref} 
\usepackage{siunitx}
\usepackage[T1]{fontenc}
\usepackage{orcidlink}

\DeclareSIUnit\clight{\text{\ensuremath{c}}}
\DeclareSIUnit\eVc{\eV/\clight} 
\DeclareSIUnit\keVc{\keV/\clight} 
\DeclareSIUnit\MeVc{\MeV/\clight} 
\DeclareSIUnit\MeVcc{\MeV/\clight\squared} 
\DeclareSIUnit\GeVc{\GeV/\clight} 
\DeclareSIUnit\GeVcc{\GeV/\clight\squared} 
\usepackage[Symbol]{upgreek}
\usepackage{textgreek}      % for greek symbols in bibliography

\begin{document}
%%%%%%%%%%%%%%%%%%%%%%%%%%%%%%%%%%%%%%%%%%%%%%%%%%
% These are some new commands that may be useful 
% for paper writing in general. If other newcommands
% are needed for your specific paper, please feel 
% free to add here. 
%
% The currently available commands are organized in: 
% 1) Systems
% 2) Quantities
% 3) Energies and units
% 4) Detectors
% 5) particle species 
%%%%%%%%%%%%%%%%%%%%%%%%%%%%%%%%%%%%%%%%%%%%%%%%%%

% 1) SYSTEMS 
\newcommand{\pp}           {pp\xspace}
\newcommand{\pbpb}           {Pb--Pb\xspace}

% 2) QUANTITIES 
\newcommand{\roots}        {\ensuremath{\sqrt{s}}\xspace}
\newcommand{\pt}           {\ensuremath{p_{\rm T}}\xspace}
\newcommand{\px}           {\ensuremath{p_{\rm x}}\xspace}
\newcommand{\py}           {\ensuremath{p_{\rm y}}\xspace}
\newcommand{\etarange}[1]  {\mbox{$\left| \eta \right| < #1$}}
\newcommand{\yrange}[1]    {\mbox{$\left| y \right| < #1$}}
\newcommand{\dndy}         {\ensuremath{\mathrm{d}N_\mathrm{ch}/\mathrm{d}y}\xspace}
\newcommand{\dndeta}       {\ensuremath{\mathrm{d}N_\mathrm{ch}/\mathrm{d}\eta}\xspace}
\newcommand{\avdndeta}     {\ensuremath{\langle\dndeta\rangle}\xspace}
\newcommand{\dNdy}         {\ensuremath{\mathrm{d}N_\mathrm{ch}/\mathrm{d}y}\xspace}
\newcommand{\dEdx}         {\ensuremath{\textrm{d}E/\textrm{d}x}\xspace}
\newcommand{\kstar}        {\ensuremath{k^*}\xspace}
\newcommand{\mt}           {\ensuremath{m_{\rm{T}}}\xspace}
\newcommand{\St}           {\ensuremath{S_{\rm{T}}}\xspace}

% 3) ENERGIES, UNITS
\newcommand{\seven}        {$\sqrt{s}~=~7$~Te\kern-.1emV\xspace}
\newcommand{\thirteen}        {$\sqrt{s}~=~13$~Te\kern-.1emV\xspace}
\newcommand{\tev}          {Te\kern-.1emV\xspace}
\newcommand{\lumi}         {\ensuremath{\mathcal{L}}\xspace}

% 4) DETECTORS 
\newcommand{\ITS}          {\rm{ITS}\xspace}
\newcommand{\TOF}          {\rm{TOF}\xspace}
\newcommand{\ZDC}          {\rm{ZDC}\xspace}
\newcommand{\ZDCs}         {\rm{ZDCs}\xspace}
\newcommand{\ZNA}          {\rm{ZNA}\xspace}
\newcommand{\ZNC}          {\rm{ZNC}\xspace}
\newcommand{\SPD}          {\rm{SPD}\xspace}
\newcommand{\SDD}          {\rm{SDD}\xspace}
\newcommand{\SSD}          {\rm{SSD}\xspace}
\newcommand{\TPC}          {\rm{TPC}\xspace}
\newcommand{\TRD}          {\rm{TRD}\xspace}
\newcommand{\VZERO}        {\rm{V0}\xspace}
\newcommand{\VZEROA}       {\rm{V0A}\xspace}
\newcommand{\VZEROC}       {\rm{V0C}\xspace}
\newcommand{\Vdecay} 	   {\ensuremath{V^{0}}\xspace}

% 4) PARTICLE SPECIES 
\newcommand{\ee}           {\ensuremath{\mathrm{e}^{+}\mathrm{e}^{-}}} 
\newcommand{\pip}          {\ensuremath{\pi^{+}}\xspace}
\newcommand{\pim}          {\ensuremath{\pi^{-}}\xspace}
\newcommand{\kap}          {\ensuremath{\mathrm{K}^{+}}\xspace}
\newcommand{\kam}          {\ensuremath{\mathrm{K}^{-}}\xspace}
\newcommand{\KK}           {\ensuremath{\mathrm{K}^{+}\mathrm{K}^{-}}\xspace}
\newcommand{\KKbar}           {\ensuremath{\mathrm{K}\overline{\mathrm{K}}}\xspace}
\newcommand{\phiPart}      {\ensuremath{\phi}\xspace}
\newcommand{\pbar}         {\ensuremath{\overline{\mathrm{p}}}\xspace}

\newcommand{\dbar}{$\overline{\mathrm{d}}$}
\newcommand{\kzero}        {\ensuremath{{\mathrm K}^{0}_{\mathrm{S}}}\xspace}
\newcommand{\lmb}          {\ensuremath{\Lambda}\xspace}
\newcommand{\almb}         {\ensuremath{\overline{\Lambda}}\xspace}
\newcommand{\Om}           {\ensuremath{\Omega^-}\xspace}
\newcommand{\Mo}           {\ensuremath{\overline{\Omega}^+}\xspace}
\newcommand{\X}            {\ensuremath{\Xi^-}\xspace}
\newcommand{\Ix}           {\ensuremath{\overline{\Xi}^+}\xspace}
\newcommand{\Xis}          {\ensuremath{\Xi^{\pm}}\xspace}
\newcommand{\Oms}          {\ensuremath{\Omega^{\pm}}\xspace}

% Femto
\newcommand{\Led}         {Lednick\'y--Lyuboshits\xspace}
\newcommand{\Ledn}         {Lednick\'y--Lyuboshits approach\xspace}
\newcommand{\Nphi}         {\ensuremath{\mathrm{N}}\mbox{--}\ensuremath{\phi}\xspace}
\newcommand{\pphi}         {\ensuremath{\mathrm{p}}\mbox{--}\ensuremath{\phi}\xspace}
\newcommand{\apphi}        {\ensuremath{\overline{\mathrm{p}}}\mbox{--}\ensuremath{\phi}\xspace}
\newcommand{\pphiComb}     {\ensuremath{\mathrm{p}}\mbox{--}\ensuremath{\phi} \ensuremath{\oplus} \ensuremath{\overline{\mathrm{p}}}\mbox{--}\ensuremath{\phi}\xspace}

\newcommand{\pd}{\ensuremath{\mbox{p--d}}~}
\newcommand{\ApAd}         {\ensuremath{\mathrm{\overline{p}}\mbox{--}\mathrm{\overline{d}}}\xspace}
\newcommand{\ld}{\ensuremath{\mbox{$\Lambda$--d}}~}

\newcommand{\pKK}         {\ensuremath{\mathrm{p}}\mbox{--}(\ensuremath{\mathrm{K}^{+}\mathrm{K}^{-}})\xspace}

\newcommand{\pP}           {\ensuremath{\mathrm{p}\mbox{--}\mathrm{p}}\xspace}
\newcommand{\pL}           {\ensuremath{\mathrm{p}\mbox{--}\Lambda}\xspace}

%\ensuremath{\mathrm\overline{p}}
\newcommand{\ApAP}         {\ensuremath{\mathrm{\overline{p}}\mbox{--}\mathrm{\overline{p}}}\xspace}

\newcommand{\pPComb}       {\ensuremath{\mathrm{p}\mbox{--}\mathrm{p} \oplus \mathrm{\overline{p}}\mbox{--}\mathrm{\overline{p}}}\xspace}

\newcommand{\pdComb}       {\ensuremath{\mathrm{p}\mbox{--}\mathrm{d} \oplus \mathrm{\overline{p}}\mbox{--}\mathrm{\overline{d}}}\xspace}

\newcommand{\pppair}           {\ensuremath{\mathrm{p}\mathrm{p}}\xspace}
\newcommand{\pdpair}           {\ensuremath{\mathrm{p}\mathrm{d}}\xspace}

\newcommand{\pppLair}           {\ensuremath{\mathrm{p}\mathrm{p}_{\Lambda}}\xspace}

\newcommand{\pLpair} {\ensuremath{\mathrm{p}\mbox{-}\Lambda}\xspace}

\newcommand{\Rpd} {\ensuremath{\mathrm{R}_\mathrm{pd}}\xspace}
\newcommand{\Rd} {\ensuremath{\mathrm{R}_\mathrm{d}}\xspace}
\newcommand{\Rp} {\ensuremath{\mathrm{R}_\mathrm{p}}\xspace}

\newcommand{\Singled} {\ensuremath{\mathrm{d}}\xspace}

% Lednicky results
\newcommand{\fReal} {\ensuremath{\Re(f_0) =\nolinebreak 0.85 \pm\nolinebreak 0.34\,(\mathrm{stat.}) \pm\nolinebreak 0.14\,(\mathrm{syst.})~\si{fm}}\xspace}
\newcommand{\fImag} {\ensuremath{\Im(f_0) =\nolinebreak 0.16 \pm\nolinebreak 0.10\,(\mathrm{stat.}) \pm\nolinebreak 0.09\,(\mathrm{syst.})~\si{fm}}\xspace}
\newcommand{\dZero} {\ensuremath{d_0 =\nolinebreak 7.85 \pm\nolinebreak 1.54\,(\mathrm{stat.})  \pm\nolinebreak 0.26\,(\mathrm{syst.})~\si{fm}}\xspace}

\newcommand{\NsigmaNoCorr} {\ensuremath{5.7 \pm\nolinebreak 0.8\,(\mathrm{stat.})  \pm\nolinebreak 0.5\,(\mathrm{syst.})~\sigma}\xspace }

\newcommand{\NsigmaNoCorrRange} {\ensuremath{4.7 - 6.6~\sigma}\xspace}

% Gaussian
\newcommand{\VeffGauss} {\ensuremath{V_{\mathrm{eff}}=\nolinebreak 2.5\pm\nolinebreak 0.9 \,(\mathrm{stat.})\pm\nolinebreak 1.4\,(\mathrm{syst.})~\si{\MeV}}\xspace}
\newcommand{\muGauss}   {\ensuremath{\upmu=\nolinebreak0.14\pm\nolinebreak 0.06 \,(\mathrm{stat.})\pm\nolinebreak 0.09\,(\mathrm{syst.})~\si{fm^{-2}}}\xspace}

% Yukawa
\newcommand{\alphaYuk}  {\ensuremath{\upalpha=\nolinebreak65.9 \pm\nolinebreak 38.0\,(\mathrm{stat.})\pm\nolinebreak 17.5\,(\mathrm{syst.})~\si{\MeV}}\xspace}
\newcommand{\AYuk}  {\ensuremath{A=\nolinebreak0.021\pm\nolinebreak 0.009\,(\mathrm{stat.})\pm\nolinebreak 0.006\,(\mathrm{syst.})}\xspace}
\newcommand{\gYuk}  {\ensuremath{g_{\Nphi} =\nolinebreak 0.14\pm\nolinebreak 0.03\,(\mathrm{stat.})\pm\nolinebreak 0.02\,(\mathrm{syst.})}\xspace}

%Normaliztion
\newcommand{\norm} {\ensuremath{\mathit{M} = 0.96}\xspace}

\newcommand{\realScatSigma}{\ensuremath{2.3\upsigma}\xspace}

%%%%%%%%%%%%%%%  Title page %%%%%%%%%%%%%%%%%%%%%%%%
\begin{titlepage}
% the dates below correspond to CERN approval
% please don't touch: EB chairs will take care
\PHyear{2024}       % required, will be obtained from CERN
\PHnumber{081}      % required, will be obtained from CERN
\PHdate{29 February}  % required, will be obtained from CERN
%%%%%%%%%%%%%%%%%%%%%%%%%%%%%%%%%%%%%%%%%%%%%%%%%%%%

\title{First measurement of the strong interaction scattering parameters for the $\mathbf{K^-d}$ and $\mathbf{K^+d}$ systems}   % appears on left page headers
\ShortTitle{Strong$-$interaction study of the kaon$-$deuteron system}   % appears on left page headers

%%% Do not change the next lines
\Collaboration{ALICE Collaboration\thanks{See Appendix~\ref{app:collab} for the list of collaboration members}}
\ShortAuthor{ALICE Collaboration} % appears on right page headers, do not change

\begin{abstract}
Charged kaon--deuteron interactions offer a sensitive probe of the strangeness sector of QCD at low energy, particularly through their scattering parameters.
Despite available theoretical predictions, experimental constraints remain absent, particularly for the $\rm K^{-}d$ system.
The first measurement of femtoscopic correlation functions for $\rm K^{-}d \oplus K^{+}\overline{d}$ and $\rm K^{+}d \oplus K^{-}\overline{d}$ particle pairs in Pb--Pb collisions at \mbox{$\sqrt{s_{\rm NN}}=5.02$~TeV} recorded by ALICE at the LHC is presented.
Correlation functions for both systems are analyzed in three centrality classes and fitted using the Lednick\'y--Lyuboshitz model to extract the source size and strong interaction scattering lengths.
The $\rm K^{-}d$ scattering length is found to be $\Re f_0 = -1.44 \pm 0.15\text{ (stat.)}^{+0.10}_{-0.10}$ (syst.) fm and $\Im f_0 = 1.34 \pm 0.33$ (stat.)$^{+0.21}_{-0.15}$ (syst.) fm, while for $\rm K^{+}d$ the value $\Re f_0 = -0.68 \pm 0.16$ (stat.)$^{+0.09}_{-0.09}$ (syst.) fm is obtained. These results provide the first constraints on the $\rm K^{\pm} d$ interaction modeling at low energy,
delivering a long-awaited experimental benchmark for testing chiral QCD dynamics.

\end{abstract}
\end{titlepage}

\setcounter{page}{2} %please do not remove this line

Studies of the interactions of matter, as embodied by nucleons ($\rm N$) and kaons ($\rm K$), the lightest pseudoscalar mesons with strangeness, play an important role in understanding quantum chromodynamics (QCD) at low energies, in the nonperturbative regime. 
This is because strange quarks, although significantly more massive than up and down quarks, are still light 
enough to justify a treatment of kaons within chiral perturbation theory, despite of the chiral symmetry breaking.
The strong interaction between the negatively charged kaon and the proton (p) has been studied through scattering experiments~\cite{PhysRev.127.1305}, kaonic atom measurements~\cite{kp12}, and femtoscopic correlation studies~\cite{Lisa2005}. 
These measurements and the corresponding theoretical calculations have established a robust foundation for studying the $\rm K^-p$ strong interaction~\cite{Martin:1980qe, Ito:1998yi, Hoshino:2017mty, Ikeda:2011pi, Ikeda:2012au, Borasoy:2006sr, Liu:2020foc, kppbpb, Meissner:2020khl, Mai:2020ltx, Hyodo:2020czb}. 
In contrast, the kaon–neutron ($\rm K^-n$) system has not been directly measured, owing to the challenges of detecting neutral particles.
However, the $\rm K^-n$ state, being a pure isospin $\rm I=1$ state, can be accessed indirectly via the $\rm K^-$ interaction with the deuteron (d), $\rm I=0,1$. Scattering in the $\rm K^-p$ and $\rm K^-d$ systems involves different linear combinations of the isospin $\rm I=0$ and $\rm I=1$ amplitudes and, therefore, measurements of the two systems allow one to disentangle the individual isospin contributions~\cite{Artibani:2024kop}. 
This is important because the $\rm I=1$ component, corresponding to $\rm K^-n$ interactions, is a key ingredient for understanding antikaons in neutron-rich matter~\cite{Tolos:2020aln, PhysRevC.90.045801, hyperonpuzzle, Gal:2016boi, Hyodo:2022xhp}.
Since the parameterization of the $\rm K^-d$ system directly impacts our understanding of the $\rm \overline{K}N$ interaction, the determination of its scattering parameters was already recognized as a priority as early as the 1980s~\cite{TOKER1981405, Deloff:2003ns} and has since been consistently emphasized by theorists as crucial for the investigation of systems with strangeness $\rm S = -1$. Furthermore, the $\rm K^-d$ interaction has been also discussed in the context of a possible quasi-bound or resonant $\rm K^-np$ state~\cite{Oset:2012gi, Shevchenko:2020kpk}, although any such state is expected to be much more weakly bound than the well-studied $\rm K^-pp$ state~\cite{Tolos:2020aln}.

So far, theoretical studies of the $\rm K^-d$ system have been performed by employing different approximations, such as the fixed-center approximation (FCA)~\cite{Kamalov:2000iy, Doring, Liu:2020foc}, or within rigorous few-body techniques~\cite{Hoshino:2017mty, Shevchenko:2014uva, Shevchenko, Mizutani, Duerinck:2025pbj}.
As input, $\rm \overline{K} N$ amplitudes derived in unitarized chiral perturbation theory or so-called "chirally motivated" phenomenological potentials have been used. 
However, no direct experimental measurements are available. Scattering experiments with low momentum kaon beams are technologically challenging and have provided only the kaon--deuteron scattering cross sections~\cite{Sibirtsev:2006yw}. 
Therefore, until now, the experimental constraints have been expected mainly from kaonic atom experiments, which can directly probe the interaction at %threshold.
zero relative momentum. 
In kaonic atom experiments, the atomic levels are modified by the presence of the strong interaction and the energy shift and width of the lowest levels are measured~\cite{kaonicatoms} through their X-ray de-excitation. 
However, the X-ray yield of kaonic deuterium is at least an order of magnitude lower than that of kaonic hydrogen, rendering such measurements exceptionally challenging~\cite{orderofmagnitude}.
Recently, the SIDDHARTA-2 experiment has significantly advanced the technology of such measurements~\cite{condmat9010016} and has completed collecting data specifically for the kaonic deuterium. 
In parallel, the femtoscopy technique has proven its ability to investigate the interaction properties of particle systems at low relative momentum, including particles with strangeness and nuclei such as deuterons~\cite{Ramos:2025ibe, PhysRevX.14.031051, wiola_pid, pdpbpb, kppbpb, kppp, ba2020135223, acharya:hal-01815196}. Therefore, the femtoscopy technique is considered as yet another powerful tool for accessing the interaction properties of the $\rm K^-d$ system.

The $\rm K^{+}d$ interaction is also highly nontrivial to study, as it cannot be accessed through kaonic atom experiments. The interaction of this system has only recently been explored in more detail for the first time using femtoscopy in pp collisions~\cite{PhysRevX.14.031051}.
The femtoscopy measurement demonstrated that the system can be effectively described using the two-body approaches anchored to scattering parameters of the KN interaction obtained through the FCA approach~\cite{Kamalov:2000iy}, based on $\rm KN$ amplitudes from chiral unitary models, or by utilizing effective range parameters deduced from explicit calculations of the $\rm K^+d$ scattering cross section with an optical potential~\cite{PhysRevX.14.031051} and in a Faddeev framework~\cite{PhysRevC.81.055204, Garcilazo:1988ma}.
However, the femtoscopic study of the $\rm K^{+}d$ system in pp collisions did not extract the scattering parameters, as it focused on particle emission properties of the system containing a deuteron.

%%%%%%%%%%%% data analysis %%%%%%%%%%%%%%%%
This study employs femtoscopy to extract the scattering lengths of kaon–deuteron systems at low relative momentum. For $\rm K^{-}d$, both the real part, reflecting the strength and sign of the elastic interaction, and the imaginary part, arising from inelastic effects, are determined. For $\rm K^{+}d$, the scattering length has only the real part, which corresponds to the elastic interaction near threshold.
The analysis is based on data collected by the ALICE experiment~\cite{production6, ALICE:2022wpn} during the Run 2 data-taking period of the Large Hadron Collider (LHC)~\cite{Evans:2008zzb}, which includes 337 million Pb--Pb collisions at the centre-of-mass energy per nucleon--nucleon pair of \mbox{$\sqrt{s_{\rm NN}}=5.02$~TeV}. 
A detailed description of the ALICE experiment can be found in Ref.~\cite{ALICE:2008ngc}. The main detectors for this study are the Inner Tracking System (ITS)~\cite{ALICE_collaboration_2010}, the Time Projection Chamber (TPC)~\cite{ALME2010316}, Time-Of-Flight (\mbox{TOF})~\cite{Akindinov:2013tea}, and two V0 detectors~\cite{Abbas:2013taa}. 

The ITS, TPC, and TOF are designed as hollow cylinders surrounding the collision point. 
They are embedded in a homogeneous magnetic field of a maximum 0.5~T provided by a solenoid magnet and are used to record particle tracks and associated information. 
The maximum pseudorapidity range shared by the three detectors is $|\eta|<0.9$. Only particles with $|\eta|<0.8$ are included in the study in order to ensure a uniform particle density.
The ITS detector, consisting of six layers of silicon detectors, is positioned closest to the beam pipe, covers the radial space from 3.9 to 85~cm, and is primarily responsible for the precise reconstruction of the collision vertex. 
Surrounding the ITS, the TPC covers the radial region from 85 cm to 247cm, and is responsible for track reconstruction, momentum determination, and particle identification.
The TPC is a gaseous detector that uses the ionisation process triggered by charged particles passing through an active volume along their trajectories.
The signals produced by traversing particles are registered in up to 159 radial rows and 18 azimuthal sectors.  
At least 70 ionization deposits associated with traversing particles are required for the reconstructed tracks to be included in this analysis.
Particle identification (PID) is performed by measuring the specific energy loss (d$E$/d$x$) in the TPC and the time-of-flight, defined as the interval between the collision time -- determined by the T0 detector~\cite{ALICE:2004ftm} -- and the detection of the signal in the TOF Multigap Resistive Plate Chambers, positioned 380 cm radially outward. 
The time-of-flight allows one to infer the mass of the particle as also the trajectory of the particle is known.
The V0 detectors are located on either side of the nominal interaction point in the beam direction with a pseudorapidity coverage of $2.8 < \eta < 5.1$ (V0A) and $-3.7 < \eta < -1.7$ (V0C) and are responsible for triggering the collision recording process. 
This process is initiated when hits are registered in both V0s in coincidence with the collision occurring in a beam bunch crossing at the centre of the ALICE detector.
The centrality of the collision expresses the percentage of the total hadronic cross section, which is proportional to the amplitudes of signals recorded in both V0s~\cite{Abelev:2013qoq}.
This work considers three centrality intervals: 0--10$\%$, 10--30$\%$, and 30--50$\%$, obtained from a minimum-bias triggered event sample~\cite{ALICEcentrality}. 
In addition, dedicated central (0--10$\%$) and semi-central (30--50$\%$) triggers are employed to enhance the statistical significance in the corresponding intervals.
To ensure a uniform acceptance, all events are required to have the collision vertex within $\pm$10~cm from the center of the ALICE detector along the beam direction.

The transverse momentum ($p_{\rm T}$) ranges of charged kaons (K$^{\pm}$) and (anti)deuterons considered in this study are 0.2–1.5~GeV$/c$ and 0.8–2.0~GeV$/c$, respectively. 
For such particles, the difference between the measured and expected signals used for PID of particles in the TPC and TOF, expressed in units of detector resolution $N\sigma$~\cite{PhysRevC.101.044907}, is calculated. The $N\sigma$ requirements depend on the particle momentum and reflect the varying level of contamination from other particle species in the detectors used for PID. For kaons, it is required that in the TPC detector $N\sigma$~<~2, 1, and 3 for momentum ($p$) ranges 0.2~<~$p$~<~0.4~GeV$/c$, 0.4~<~$p$~<~0.45~GeV$/c$, and 0.45~<~$p$~<~1.5~GeV$/c$, respectively, while in the TOF detector $N\sigma$~<~2, 1.5, 1 is required for 0.45~<~$p$~<~0.8~GeV$/c$, 0.8~<~$p$~<~1.0~GeV$/c$, and 1~<~$p$~<~1.5~GeV$/c$, respectively. 
For (anti)deuterons, $N\sigma$~<~2 is employed in the full momentum range for the TPC, with an additional TOF $N\sigma$ < 2 requirement for 1.3~<~$p$~<~2.0~GeV$/c$.
Additionally, the distance of closest approach (DCA) of the particle track to the collision vertex for kaons and antideuterons is required to be smaller than 2.4~cm and 3.2~cm in the transverse and beam directions, respectively. 
Deuterons are expected to be subject to non-primary contamination from spallation processes, with these contributions generally directed away from the collision vertex.
To reduce the non-primary contribution, the deuteron's DCA is limited to a maximum $0.0105 + 0.0350/(p_{\text{T}}/({\rm{GeV}}/c))^{1.1} ~\text{cm}$ in the transverse plane and 1 cm in the beam direction. 

In addition, similar to the approach adopted in Refs.~\cite{kppbpb, wiola_pid, kppp, pdpbpb}, particle pairs are subjected to selection criteria designed to reduce two-particle effects, which come primarily from detector acceptance, track merging, and contamination from photon conversions.
The effect from photon conversions is removed by requiring the invariant mass of the pair, calculated assuming both tracks are electrons, to be larger than 0.002 GeV$/c^2$, and the relative polar angle between the two tracks to be larger than 0.008 rad. The points of two tracks in the TPC are considered as merged if the distance between them is less than 3 cm. This distance is evaluated for pairs of tracks with a relative pseudorapidity difference of less than 0.01 with a step of 1 cm along the TPC radius. Track pairs with a merged point fraction exceeding 2$\%$ are removed.

The correlation function is measured as $C_{\text{exp}}({k}^*) = \EuScript{N}A(k^*)/B(k^*)$, where $A$ and $B$ correspond to the correlated and uncorrelated distributions of $k^*$, which is the half of the pair's relative momentum in the pair rest frame (PRF -- when the pair's centre of mass is at rest), respectively, and $\EuScript{N}$ is a normalization factor. 
The physically correlated distribution is constructed from particle pairs originating from the same collision, while the uncorrelated distribution (without physical interactions) is constructed by creating pairs of particles originating from different collisions with similar collision vertex position and centrality.
The $\EuScript{N}$ factor is determined in the $k^*$ region where femtoscopic and non-femtoscopic correlations are expected to be negligible -- chosen in this work to be 0.25--0.50 GeV/$c$.
The function $C$ ($k^*$) is equal to unity when there is no correlation.
A deviation above (below) unity indicates an attractive (repulsive) interaction, while a bound state manifests as a systematic and distinctive depletion in the correlation function at a $k^*$ value which depends on the binding energy~\cite{Acharya:2020asf, oton}.
A more localized area of particle emission increases the contribution of close-distance particle pairs with pronounced interaction effects in the correlation function. Thus, the strength of the measured correlation function for the same particle pairs vary at different collision energies and types. In Pb$-$Pb collisions, the contribution from pairs with large relative distances is greater, causing the interaction effects to manifest mainly in the very low-$k^*$ region, which in this work is $k^*<0.20~{\rm{GeV}}/c$. Nevertheless, correlation functions can be studied differentially across collision centralities, and thus multiplicities, which provides a laboratory for studying interaction effects in different environments of particle emission~\cite{kppbpb, ppbramona, ba2020135223}.

The theoretical correlation function of particle pairs at low relative momentum can be expressed via the Koonin--Pratt formula~\cite{Koonin:1977fh,Pratt:1990zq}
\begin{equation}
     C(\boldsymbol{k}^*) = \int S(r^*) |\Psi(r^{*},\boldsymbol{k}^*)|^2 \text{d}^3 r^* \, \label{eq:kooninpratt}
\end{equation}
where the variables are defined in the PRF, $\boldsymbol{k}^* $ is half of the pair's relative momentum vector, and $\boldsymbol{r}^* $ is the relative distance vector. In this study, the source function $S(\boldsymbol{r}^*)$ describes the spatial emission properties of the two particles~\cite{sinyukovexpanding}. This function can be parameterized by a Gaussian, with the standard deviation interpreted as the femtoscopic source radius $R_{\rm Kd}$ for the Kd pairs~\cite{Kisiel:2009eh}. 
The pair wave function $\Psi(\boldsymbol{r}^*,\boldsymbol{k}^*)$ encodes the Coulomb and strong interactions between kaons and deuterons.
Since in the PRF particles are moving slowly with respect to each other, minimizing the effect of non-equal emission time for $k^{*}$ $\rightarrow 0$, the wave function can be described by the Bethe--Salpeter amplitude in the Lednický--Lyuboshitz formalism ~\cite{Lednicky:1981su, Lednicky:2003mq, Lednicky:2009zza} (Supplemental Material~\ref{sec:appendix:LL}). 
In this work, the interaction parameters such as the scattering lengths ($f_{\rm{0}}$) are obtained, while the effective range ($d_{\rm 0}$) of the interaction is approximated to zero~\cite{kppbpb}.

This femtoscopic analysis is conducted for $\rm K^{-}d$ and $\rm K^{+}d$ pairs and their charge conjugates in ultrarelativistic heavy-ion collisions. 
The experimental correlation function may contain non-femtoscopic effects, e.g., those originating from the collective motion of the system~\cite{Kisiel:2017gip}.
The impact of these effects can be estimated by fitting the correlation function with a first-order polynomial in the normalization region and extrapolating the resulting baseline to $k^*$ = 0 ($C_{\text{bg}}$). To remove non-femtoscopic effects, the raw correlation function is then divided by such extrapolation. 
The correlation functions are also subject to corrections which account for imperfect particle identification, contamination from non-primary particles, and non-gaussianity of the source. 
The fractions of pairs affected by the first two effects are calculated for kaons and (anti)deuterons based on dedicated Monte Carlo (MC) simulations and a data-driven approach as described in Refs.~\cite{kppbpb, acharya:hal-01815196}. 
The Gaussianity correction factor of the source is calculated following the procedure in Ref.~\cite{Kisiel:2009eh}. 
The total correction factor ($f_{\text{corr}}$) is evaluated as a function of $k^*$. Averaged over the low-$k^*$ region, the total correction factor is found to be 0.70, 0.75, and 0.80 for $\rm K^{+}d$ and $\rm K^{-}d$ pairs, and 0.84, 0.85, and 0.87 for $\rm K^{+}\overline{d}$ and $\rm K^{-}\overline{d}$ in the 0--10\%, 10--30\%, and 30--50\% centrality intervals, respectively.
The aforementioned corrections to the experimental functions are implemented in accordance with the formula 
\begin{equation}
C(k^*) = \left( \frac{C_{\text{exp}}(k^*) - 1}{f_{\text{corr}}(k^*)} + 1 \right) / C_{\text{bg}}(k^*).
    \label{eq:corr}
\end{equation} 

\begin{figure}[h!]
    \centering
    \includegraphics[width=1.0\textwidth]{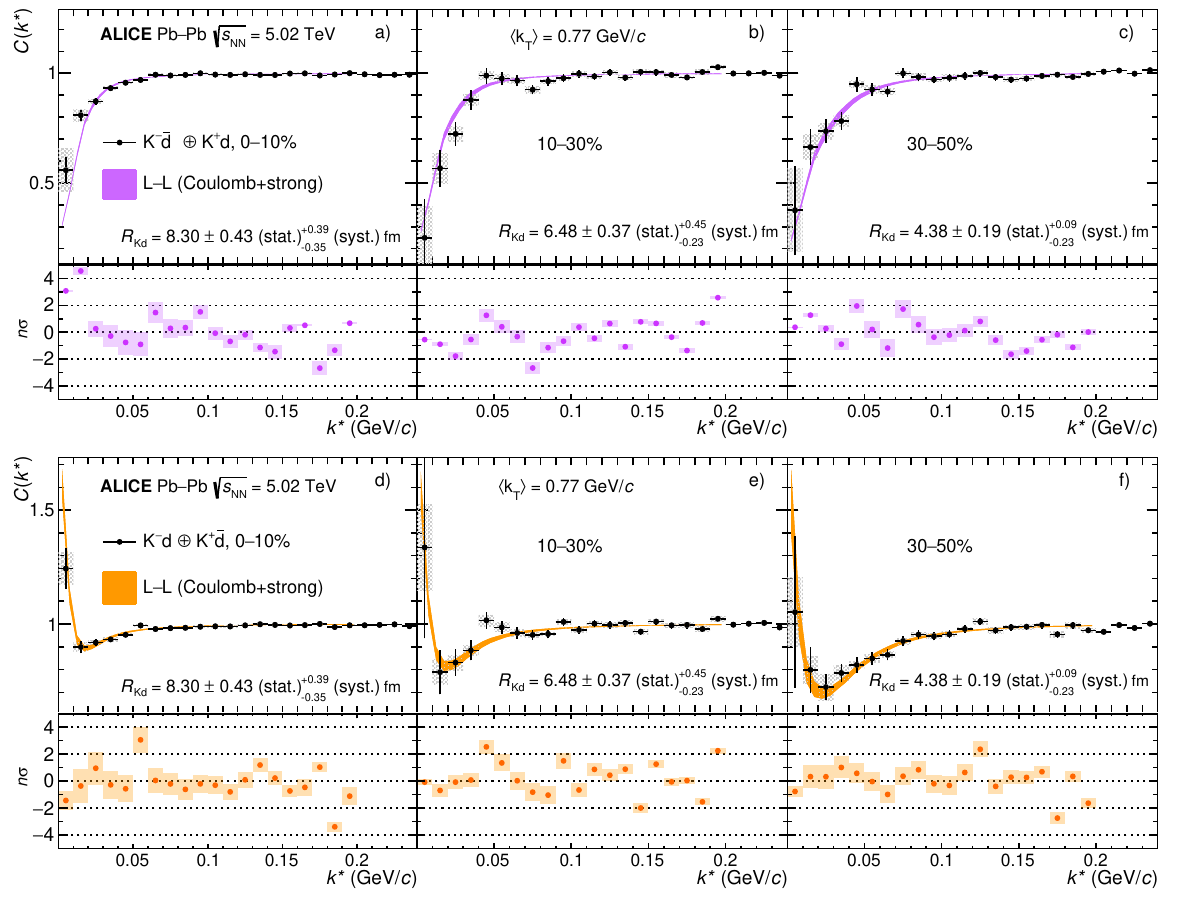}
    \caption{The $\rm K^{+}d$ $\oplus$ $\rm K^{-}\overline{d}$ (top pads of panels a, b, c), and  $\rm K^{-}d$ $\oplus$ $\rm K^{+}\overline{d}$ (top pads of panels d, e, f), correlation functions in the three centrality classes 0$-$10$\%$, 10$-$30$\%$, 30$-$50$\%$ represented with markers. The vertical lines and gray boxes correspond to the statistical and systematic uncertainties, respectively. The light violet and orange bands corresponds to Lednick\'{y}--Lyuboshits fits (denoted as L--L), including Coulomb and strong interactions. The width of the bands represents the systematic uncertainties of the fit. The bottom pads in all panels show the difference between the data and the L--L fit normalised by the statistical uncertainty of the data $\sigma_{\text{stat}}$.}
    \label{fig:monsterplot_kd}
\end{figure}

Since pairs and their charge conjugates are expected to be governed by the same interactions and no systematic deviations between their correlation functions are observed, they are combined together after corrections.
The experimental correlation function obtained using Eq.~(\ref{eq:corr})
is presented in Fig.~\ref{fig:monsterplot_kd}. 
The effects of Coulomb and strong interactions manifest themselves in the low-$k^{*}$ range of the $\rm K^{+}d\oplus K^{-}\overline{d}$ and $\rm K^{-}d\oplus K^{+}\overline{d}$ systems as an enhancement above unity and a depletion below unity, respectively.
The correlation function of kaon--deuteron pairs behaves similarly to kaon--proton correlation functions~\cite{kppbpb}.

The correlation functions are presented together with statistical and systematic uncertainties in Fig.~\ref{fig:monsterplot_kd}. 
Systematic uncertainties include variations of event and particle selection criteria.
These include varying the pseudorapidity range by $\pm$0.03 and changing the minimum number of clusters in TPC required for track reconstruction by $\pm$10.
Also, the selection of the position of the primary vertex is varied to be within $\pm$8~cm. The PID selection criteria, $N\sigma$, are adjusted by a change of $\pm$0.5$\sigma$. Furthermore, the strict requirement on the deuteron's DCA in the transverse plane is relaxed to 0.1~cm. The merged fraction in the pair-track selections is varied by $\pm$1$\%$. The systematic uncertainty of each $k^*$ point of the correlation function shown in Fig.~\ref{fig:monsterplot_kd} reflects the one standard deviation resulting from the combination of all the above-mentioned variations of the selection criteria. 

The correlation functions shown in Fig.~\ref{fig:monsterplot_kd} are calculated using the Lednick\'{y}--Lyuboshits formalism with inclusion of the Coulomb interaction~\cite{Lednicky:1981su, kppbpb} based on the values of the source size ($R_{\rm Kd})$ and scattering parameters ($\Re f_{\rm 0}^{\rm K^{-}d}$, $\Im f_{\rm 0}^{\rm K^{-}d}$, $\Re f_{\rm 0}^{\rm K^{+}d}$) obtained in the fit to the data. 
The presence of the Coulomb interaction requires the inclusion
of its effects in the employed formulae and the resulting scattering parameters describe Coulomb-modified quantities. 
Such model correlation functions are corrected for detector momentum resolution effects using the detector response matrix obtained from the MC HIJING~\cite{Wang:1991hta} simulation coupled to the GEANT3~\cite{Brun:1994aa} transport code. 
All six correlation functions are simultaneously fitted in the low-$k^*$ region as previously done in Refs.~\cite{kppbpb, wiola_pid}, such that the source radius parameters are shared among centralities for same- and opposite-sign pairs, while the scattering parameters are shared across different centralities for same/opposite-sign pairs.
The n-$\sigma$ distribution in the lower panel shows that the fit accurately describes the data for all pairs across the three centrality classes considered in the study.
Systematic uncertainties of the fit parameters are evaluated from the spread of results obtained with several systematic variations. 
These include: statistical resampling of the correlation function space points; modifications of the correlation function or the fitting procedure (with non-femtoscopic backgrounds described by first- and second-order polynomials in the ranges 0.30--0.55~GeV$/c$ and 0.20--0.90~GeV$/c$, respectively); variation of the total correction factor by $\pm$ 0.05 to account for a possible under- or overestimation of PID purities and the primary contributors in the particle sample; changes of the fit range in $k^*$ between 0.15 and 0.25 GeV/$c$; and simultaneous fits constrained by theoretical predictions for the $\rm K^+d$ scattering length using the mentioned theoretical predictions from the FCA and ER calculations. 
The uncertainty band in Fig.~\ref{fig:monsterplot_kd} corresponds to one standard deviation of the fit solutions across all such variations.

The obtained two-particle radius $R_{\rm K d}$ increases from $4.39 \pm 0.18$ (stat.)$^{+0.09}_{-0.23}$ (syst.)~fm for 30--50$\%$ centrality, through $6.48 \pm 0.37$ (stat.)$^{+0.45}_{-0.23}$ (syst.)~fm for 10--30$\%$ to $8.30 \pm 0.43$ (stat.)$^{+0.39}_{-0.35}$ (syst.)~fm for 0--10$\%$ centrality. The increase of the femtoscopic radius with increasing centrality is expected and observed in other studies~\cite{therm2014,therm2018, ALICE:2015hvw}, as the system size increases with particle multiplicity in more central collisions. 

\begin{figure}[htb!]
    \centering
    \includegraphics[width=0.72\textwidth]{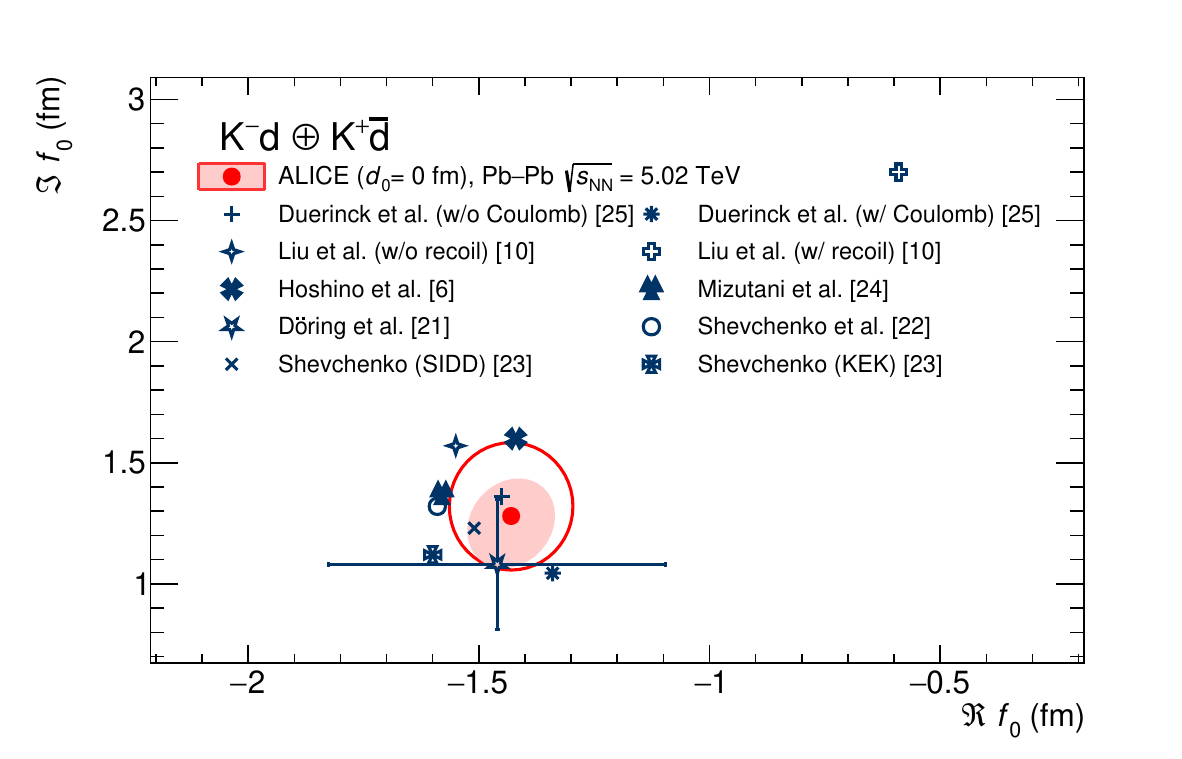}
    \caption{Scattering lengths of $\rm K^{-}d$: Lednick\'{y}--Lyuboshits fit obtained in this study (red marker), available theoretical calculations using input data from SIDDHARTA and KEK experiments (blue markers)~\cite{Duerinck:2025pbj, Liu:2020foc, Hoshino:2017mty, Mizutani, Shevchenko:2014uva, Shevchenko, Doring}. The full and empty ellipses correspond to systematic and statistical uncertainties of the ALICE measurement, respectively.}
    \label{fig:scat_kd1}
\end{figure}
\begin{figure}[htb!]
    \centering
    \includegraphics[width=0.72\textwidth]{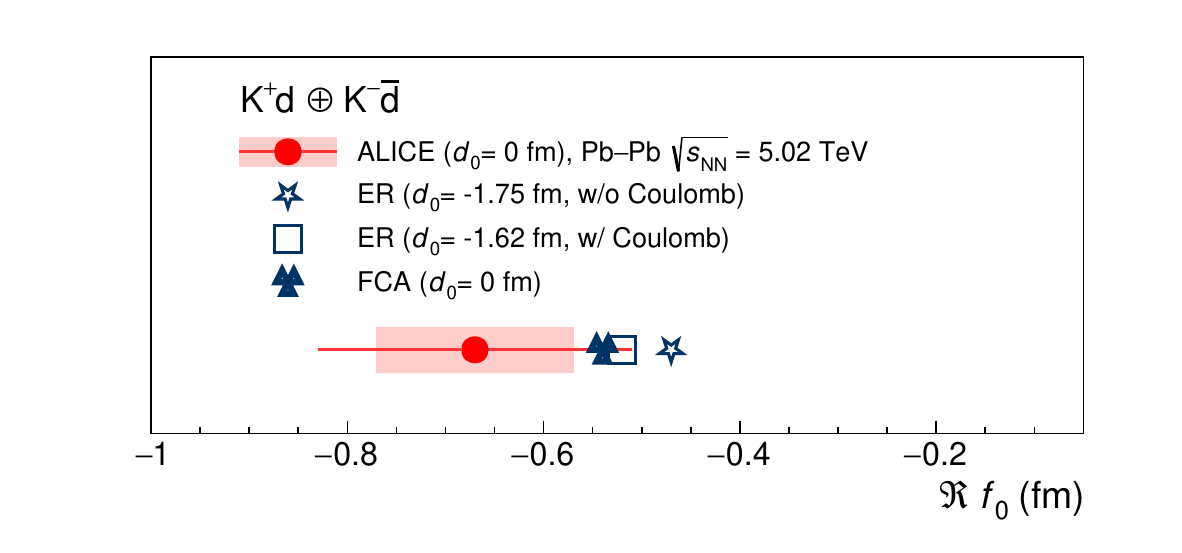 }   
    \caption{Scattering lengths of $\rm K^{+}d$: Lednick\'{y}--Lyuboshits fit obtained in this study (red marker), available theoretical calculations (blue marker) provided with the FCA~\cite{Kamalov:2000iy} with the KN scattering lengths from Ref.~\cite{Aoki:2018wug} and effective range fit (ER)~\cite{PhysRevC.81.055204} to the cross section predictions at threshold anchored to the scattering data~\cite{PhysRevC.81.055204}. The full box and line correspond to systematic and statistical uncertainties of the ALICE measurement, respectively.}
    \label{fig:scat_kd2}
\end{figure}

The obtained values of the Coulomb modified $\rm K^{-}d$ scattering length are $\Re f_{\rm 0}^{\rm K^{-}d}=$ $-1.44 \pm 0.15$ (stat.)$^{+0.10}_{-0.10}$ (syst.) fm and \mbox{$\Im f_{\rm 0}^{\rm K^{-}d} =$ $1.34 \pm 0.33$ (stat.)$^{+0.21}_{-0.15}$ (syst.) fm}. In Fig.~\ref{fig:scat_kd1}, they are compared with recent model calculations~\cite{Doring, Hoshino:2017mty, Mizutani, Shevchenko, Shevchenko:2014uva, Liu:2020foc, Duerinck:2025pbj}. Most theoretical predictions which are based on kaonic hydrogen data from the SIDDHARTA~\cite{Zmeskal:2009zz} and KEK~\cite{kek} experiments are consistent with the ALICE data, though the extracted parameters favour models that do not incorporate an additional recoil effect on the $\rm \overline{K}N$ energy, as discussed in Ref.~\cite{Liu:2020foc}. However, it should be said that Coulomb effects are not included in the calculations, except for Ref.~\cite{Duerinck:2025pbj}.

In the case of the $\rm K^{+}d$ system, the obtained Coulomb modified scattering length value is $\Re f_{\rm 0}^{\rm K^{+}d} = -0.68 \pm 0.16$ (stat.)$^{+0.09}_{-0.09}$ (syst.)~fm. As shown in Fig.~\ref{fig:scat_kd2}, the measurement is in good agreement with both theoretical calculations employing FCA and ER approaches, which were previously used to successfully describe the femtoscopic correlation function of the $\rm K^{+}d$ system in pp collisions~\cite{PhysRevX.14.031051}.

%----------------------summary------------------------
The determination of the K$^{-}$d and K$^{+}$d scattering lengths presented here constitutes the first direct experimental access to the strong interaction in the kaon–deuteron system. Unlike kaonic-atom studies and traditional scattering experiments, this femtoscopic approach provides measurements that are free from extrapolation to threshold and avoids the need to translate level shifts and widths into scattering lengths. The extracted K$^{-}$d, K$^{+}$d values are consistent with recent theoretical calculations and can improve our knowledge of the low-energy $\mathrm{K^{-}n}$ interaction, which lacks direct experimental constraints. Since the deuteron contains both proton and neutron components, the measured scattering lengths probe a mixture of the $\mathrm{I=0}$ and $\mathrm{I=1}$ kaon–nucleon amplitudes, thereby providing complementary information to single-nucleon systems. Together with existing measurements of the $\mathrm{K^{\pm}p}$ interactions, these results establish a comprehensive experimental benchmark for QCD-based chiral models and enable an improved characterization of the $\mathrm{K^{\pm}N}$ sector.

%%%%%%%%%%%%%%%%%%%%%%%%%%%%%%%%
% end main text 
%%%%%%%%%%%%%%%%%%%%%%%%%%%%%%%%

%%%%% acknowledgements - handled by EB chairs 
\newenvironment{acknowledgement}{\relax}{\relax}
\begin{acknowledgement}
\section*{Acknowledgements}
% add specific acknowledgements here 
% ...but please don't remove the line below: funding agencies
% will be acknowledged with a custom tex file handled by EB chairs after Collab Round 2
% Version: 2026-02-04

The ALICE Collaboration would like to thank all its engineers and technicians for their invaluable contributions to the construction of the experiment and the CERN accelerator teams for the outstanding performance of the LHC complex.
The ALICE Collaboration gratefully acknowledges the resources and support provided by all Grid centres and the Worldwide LHC Computing Grid (WLCG) collaboration.
The ALICE Collaboration acknowledges the following funding agencies for their support in building and running the ALICE detector:
A. I. Alikhanyan National Science Laboratory (Yerevan Physics Institute) Foundation (ANSL), State Committee of Science and World Federation of Scientists (WFS), Armenia;
Austrian Academy of Sciences, Austrian Science Fund (FWF): [M 2467-N36] and Nationalstiftung f\"{u}r Forschung, Technologie und Entwicklung, Austria;
Ministry of Communications and High Technologies, National Nuclear Research Center, Azerbaijan;
Rede Nacional de Física de Altas Energias (Renafae), Financiadora de Estudos e Projetos (Finep), Funda\c{c}\~{a}o de Amparo \`{a} Pesquisa do Estado de S\~{a}o Paulo (FAPESP) and The Sao Paulo Research Foundation  (FAPESP), Brazil;
Bulgarian Ministry of Education and Science, within the National Roadmap for Research Infrastructures 2020-2027 (object CERN), Bulgaria;
Ministry of Education of China (MOEC) , Ministry of Science \& Technology of China (MSTC) and National Natural Science Foundation of China (NSFC), China;
Ministry of Science and Education and Croatian Science Foundation, Croatia;
Centro de Aplicaciones Tecnol\'{o}gicas y Desarrollo Nuclear (CEADEN), Cubaenerg\'{\i}a, Cuba;
Ministry of Education, Youth and Sports of the Czech Republic, Czech Republic;
The Danish Council for Independent Research | Natural Sciences, the VILLUM FONDEN and Danish National Research Foundation (DNRF), Denmark;
Helsinki Institute of Physics (HIP), Finland;
Commissariat \`{a} l'Energie Atomique (CEA) and Institut National de Physique Nucl\'{e}aire et de Physique des Particules (IN2P3) and Centre National de la Recherche Scientifique (CNRS), France;
Bundesministerium f\"{u}r Forschung, Technologie und Raumfahrt (BMFTR) and GSI Helmholtzzentrum f\"{u}r Schwerionenforschung GmbH, Germany;
National Research, Development and Innovation Office, Hungary;
Department of Atomic Energy Government of India (DAE), Department of Science and Technology, Government of India (DST), University Grants Commission, Government of India (UGC) and Council of Scientific and Industrial Research (CSIR), India;
National Research and Innovation Agency - BRIN, Indonesia;
Istituto Nazionale di Fisica Nucleare (INFN), Italy;
Japanese Ministry of Education, Culture, Sports, Science and Technology (MEXT) and Japan Society for the Promotion of Science (JSPS) KAKENHI, Japan;
Consejo Nacional de Ciencia (CONACYT) y Tecnolog\'{i}a, through Fondo de Cooperaci\'{o}n Internacional en Ciencia y Tecnolog\'{i}a (FONCICYT) and Direcci\'{o}n General de Asuntos del Personal Academico (DGAPA), Mexico;
Nederlandse Organisatie voor Wetenschappelijk Onderzoek (NWO), Netherlands;
The Research Council of Norway, Norway;
Pontificia Universidad Cat\'{o}lica del Per\'{u}, Peru;
Ministry of Science and Higher Education, National Science Centre and WUT ID-UB, Poland;
Korea Institute of Science and Technology Information and National Research Foundation of Korea (NRF), Republic of Korea;
Ministry of Education and Scientific Research, Institute of Atomic Physics, Ministry of Research and Innovation and Institute of Atomic Physics and Universitatea Nationala de Stiinta si Tehnologie Politehnica Bucuresti, Romania;
Ministerstvo skolstva, vyskumu, vyvoja a mladeze SR, Slovakia;
National Research Foundation of South Africa, South Africa;
Swedish Research Council (VR) and Knut \& Alice Wallenberg Foundation (KAW), Sweden;
European Organization for Nuclear Research, Switzerland;
Suranaree University of Technology (SUT), National Science and Technology Development Agency (NSTDA) and National Science, Research and Innovation Fund (NSRF via PMU-B B05F650021), Thailand;
Turkish Energy, Nuclear and Mineral Research Agency (TENMAK), Turkey;
National Academy of  Sciences of Ukraine, Ukraine;
Science and Technology Facilities Council (STFC), United Kingdom;
National Science Foundation of the United States of America (NSF) and United States Department of Energy, Office of Nuclear Physics (DOE NP), United States of America.
In addition, individual groups or members have received support from:
FORTE project, reg.\ no.\ CZ.02.01.01/00/22\_008/0004632, Czech Republic, co-funded by the European Union, Czech Republic;
European Research Council (grant no. 950692), European Union;
Deutsche Forschungs Gemeinschaft (DFG, German Research Foundation) ``Neutrinos and Dark Matter in Astro- and Particle Physics'' (grant no. SFB 1258), Germany;
FAIR - Future Artificial Intelligence Research, funded by the NextGenerationEU program (Italy).

\end{acknowledgement}

%%%%%%%% Bibliography 
\bibliographystyle{utphys}   % Remember we use title in the biblio
\bibliography{apssamp}

\newpage
\appendix

\section{Lednick\'y--Lyuboshitz formalism}
\label{sec:appendix:LL}
The pair wave function $\Psi(\boldsymbol{k}^{},\boldsymbol{r}^{})$ in Eq.~(\ref{eq:kooninpratt}) reflects the underlying two-particle interaction. For the case of charged kaons scattering off deuterons, both electromagnetic and strong forces contribute. For such non-identical particle pairs, the interaction is represented through the Bethe–Salpeter amplitude, which is the solution to the corresponding quantum scattering problem approached in reverse time (indicated by the ``$-$'' sign preceding $k^*$)
\vspace{0.25cm}\\
\begin{equation}
\Psi^{(+)}_{-{\boldsymbol{k}^{*}}}({\boldsymbol{r}^{*}}) = \sqrt{A_{\rm C} (\varepsilon)}
\frac{1}{\sqrt{2}} \left [ {\rm e}^{-{\rm i} \boldsymbol{k}^{*}\cdot{\boldsymbol{r}^{*}}} {\rm F}(-{\rm i} \varepsilon, 1,
  {\rm i} \zeta^{+}) + f_{\rm C}(\boldsymbol{k}^*)\frac{\tilde{G}(\rho,\varepsilon)}{{r}^*} \right ],
\label{eq:fullpsi2}
\end{equation}
\vspace{0.01cm}\\
where $A_{\rm C}$ is the Gamow factor,
$\varepsilon = 1/(k^{*} a_{\rm C})$, 
$\zeta^{\pm} = k^{*} r^{*} (1 \pm \cos{\theta^{*}})$, 
$\rm F$ is the confluent hypergeometric function, and $\tilde{G}$ represents a specific combination of the regular and singular s-wave Coulomb functions.
The angle $\theta^{*}$ specifies the orientation between the relative momentum $k^*$ and the relative separation $r^{*}$, whereas $a_{\rm C}$ represents the Bohr radius of the pair. The term $f_{\rm C}(k^*)$ corresponds to the strong-interaction scattering amplitude, modified to consistently include the effect of the Coulomb interaction.

\begin{equation}
\label{eq:scattering_amplitude}
    f^{-1}_C(k^*) = \frac{1}{f_0} + \frac{1}{2}d_0{k^*}^2-\frac{2}{a_C}h(k^* a_C) -\text{i}k^* a_C,
\end{equation}
where $h(\varepsilon)=\varepsilon^2\sum\limits_{n=1}^{ \infty}[n(n^2+\varepsilon^2)]^{-1}
-\mathrm{{\gamma}} -\ln|\varepsilon|$
($\mathrm{{\gamma}} = 0.5772$ represents the Euler constant).

The correlation function is obtained by numerically integrating the source function $S(r^*)$, modeled as a three-dimensional Gaussian in the PRF, where the particle momenta satisfy ($\boldsymbol{p_1}^*=-\boldsymbol{p_2}^*$). This integration is carried out using the Bethe–Salpeter amplitude given in Eq.~(\ref{eq:fullpsi2}), together with the Coulomb-modified scattering amplitude introduced in Eq.~(\ref{eq:scattering_amplitude})~\cite{Kisiel:2009eh,Kisiel:2018wie}.

\clearpage
%%%%% Authorlist - please do not touch: handled by EB chairs 
\section{The ALICE Collaboration}
\label{app:collab}
% ALICE Collaboration author list for 2026-02-04
\begin{flushleft} 
\small

D.A.H.~Abdallah\,\orcidlink{0000-0003-4768-2718}\,$^{\rm 136}$, 
I.J.~Abualrob\,\orcidlink{0009-0005-3519-5631}\,$^{\rm 114}$, 
S.~Acharya\,\orcidlink{0000-0002-9213-5329}\,$^{\rm 49}$, 
K.~Agarwal\,\orcidlink{0000-0001-5781-3393}\,$^{\rm II,}$$^{\rm 23}$, 
G.~Aglieri Rinella\,\orcidlink{0000-0002-9611-3696}\,$^{\rm 32}$, 
L.~Aglietta\,\orcidlink{0009-0003-0763-6802}\,$^{\rm 24}$, 
N.~Agrawal\,\orcidlink{0000-0003-0348-9836}\,$^{\rm 25}$, 
Z.~Ahammed\,\orcidlink{0000-0001-5241-7412}\,$^{\rm 134}$, 
S.~Ahmad\,\orcidlink{0000-0003-0497-5705}\,$^{\rm 15}$, 
I.~Ahuja\,\orcidlink{0000-0002-4417-1392}\,$^{\rm 36}$
% ...

\section*{Affiliation Notes}

$^{\rm I}$ Deceased\\
$^{\rm II}$ Also at: INFN Trieste\\
$^{\rm III}$ Also at: Max-Planck-Institut fur Physik, Munich, Germany\\
$^{\rm IV}$ Also at: Czech Technical University in Prague (CZ)\\
$^{\rm V}$ Also at: Instituto de Fisica da Universidade de Sao Paulo\\
$^{\rm VI}$ Also at: Dipartimento DET del Politecnico di Torino, Turin, Italy\\
$^{\rm VII}$ Also at: Department of Applied Physics, Aligarh Muslim University, Aligarh, India\\
$^{\rm VIII}$ Also at: Institute of Theoretical Physics, University of Wroclaw, Poland\\
$^{\rm IX}$ Also at: Facultad de Ciencias, Universidad Nacional Aut\'{o}noma de M\'{e}xico, Mexico City, Mexico\\

\section*{Collaboration Institutes}

$^{1}$ A.I. Alikhanyan National Science Laboratory (Yerevan Physics Institute) Foundation, Yerevan, Armenia\\
$^{2}$ AGH University of Krakow, Cracow, Poland\\
$^{3}$ Bogolyubov Institute for Theoretical Physics, National Academy of Sciences of Ukraine, Kyiv, Ukraine\\
$^{4}$ Bose Institute, Department of Physics  and Centre for Astroparticle Physics and Space Science (CAPSS), Kolkata, India\\
$^{5}$ California Polytechnic State University, San Luis Obispo, California, United States\\
$^{6}$ Central China Normal University, Wuhan, China\\
$^{7}$ Centro de Aplicaciones Tecnol\'{o}gicas y Desarrollo Nuclear (CEADEN), Havana, Cuba\\
$^{8}$ Centro de Investigaci\'{o}n y de Estudios Avanzados (CINVESTAV), Mexico City and M\'{e}rida, Mexico\\
$^{9}$ Chicago State University, Chicago, Illinois, United States\\
$^{10}$ China Nuclear Data Center, China Institute of Atomic Energy, Beijing, China\\
$^{11}$ China University of Geosciences, Wuhan, China\\
$^{12}$ Chungbuk National University, Cheongju, Republic of Korea\\
$^{13}$ Comenius University Bratislava, Faculty of Mathematics, Physics and Informatics, Bratislava, Slovak Republic\\
$^{14}$ Creighton University, Omaha, Nebraska, United States\\
$^{15}$ Department of Physics, Aligarh Muslim University, Aligarh, India\\
$^{16}$ Department of Physics, Pusan National University, Pusan, Republic of Korea\\
$^{17}$ Department of Physics, Sejong University, Seoul, Republic of Korea\\
$^{18}$ Department of Physics, University of California, Berkeley, California, United States\\
$^{19}$ Department of Physics, University of Oslo, Oslo, Norway\\
$^{20}$ Department of Physics and Technology, University of Bergen, Bergen, Norway\\
$^{21}$ Dipartimento di Fisica, Universit\`{a} di Pavia, Pavia, Italy\\
$^{22}$ Dipartimento di Fisica dell'Universit\`{a} and Sezione INFN, Cagliari, Italy\\
$^{23}$ Dipartimento di Fisica dell'Universit\`{a} and Sezione INFN, Trieste, Italy\\
$^{24}$ Dipartimento di Fisica dell'Universit\`{a} and Sezione INFN, Turin, Italy\\
$^{25}$ Dipartimento di Fisica e Astronomia dell'Universit\`{a} and Sezione INFN, Bologna, Italy\\
$^{26}$ Dipartimento di Fisica e Astronomia dell'Universit\`{a} and Sezione INFN, Catania, Italy\\
$^{27}$ Dipartimento di Fisica e Astronomia dell'Universit\`{a} and Sezione INFN, Padova, Italy\\
$^{28}$ Dipartimento di Fisica `E.R.~Caianiello' dell'Universit\`{a} and Gruppo Collegato INFN, Salerno, Italy\\
$^{29}$ Dipartimento DISAT del Politecnico and Sezione INFN, Turin, Italy\\
$^{30}$ Dipartimento di Scienze MIFT, Universit\`{a} di Messina, Messina, Italy\\
$^{31}$ Dipartimento Interateneo di Fisica `M.~Merlin' and Sezione INFN, Bari, Italy\\
$^{32}$ European Organization for Nuclear Research (CERN), Geneva, Switzerland\\
$^{33}$ Faculty of Electrical Engineering, Mechanical Engineering and Naval Architecture, University of Split, Split, Croatia\\
$^{34}$ Faculty of Nuclear Sciences and Physical Engineering, Czech Technical University in Prague, Prague, Czech Republic\\
$^{35}$ Faculty of Physics, Sofia University, Sofia, Bulgaria\\
$^{36}$ Faculty of Science, P.J.~\v{S}af\'{a}rik University, Ko\v{s}ice, Slovak Republic\\
$^{37}$ Faculty of Technology, Environmental and Social Sciences, Bergen, Norway\\
$^{38}$ Frankfurt Institute for Advanced Studies, Johann Wolfgang Goethe-Universit\"{a}t Frankfurt, Frankfurt, Germany\\
$^{39}$ Fudan University, Shanghai, China\\
$^{40}$ Gauhati University, Department of Physics, Guwahati, India\\
$^{41}$ Helmholtz-Institut f\"{u}r Strahlen- und Kernphysik, Rheinische Friedrich-Wilhelms-Universit\"{a}t Bonn, Bonn, Germany\\
$^{42}$ Helsinki Institute of Physics (HIP), Helsinki, Finland\\
$^{43}$ High Energy Physics Group,  Universidad Aut\'{o}noma de Puebla, Puebla, Mexico\\
$^{44}$ Horia Hulubei National Institute of Physics and Nuclear Engineering, Bucharest, Romania\\
$^{45}$ HUN-REN Wigner Research Centre for Physics, Budapest, Hungary\\
$^{46}$ Indian Institute of Technology Bombay (IIT), Mumbai, India\\
$^{47}$ Indian Institute of Technology Indore, Indore, India\\
$^{48}$ INFN, Laboratori Nazionali di Frascati, Frascati, Italy\\
$^{49}$ INFN, Sezione di Bari, Bari, Italy\\
$^{50}$ INFN, Sezione di Bologna, Bologna, Italy\\
$^{51}$ INFN, Sezione di Cagliari, Cagliari, Italy\\
$^{52}$ INFN, Sezione di Catania, Catania, Italy\\
$^{53}$ INFN, Sezione di Padova, Padova, Italy\\
$^{54}$ INFN, Sezione di Pavia, Pavia, Italy\\
$^{55}$ INFN, Sezione di Torino, Turin, Italy\\
$^{56}$ INFN, Sezione di Trieste, Trieste, Italy\\
$^{57}$ Inha University, Incheon, Republic of Korea\\
$^{58}$ Institute for Advanced Simulation, Forschungszentrum J\"ulich, J\"ulich, Germany\\
$^{59}$ Institute for Gravitational and Subatomic Physics (GRASP), Utrecht University/Nikhef, Utrecht, Netherlands\\
$^{60}$ Institute of Experimental Physics, Slovak Academy of Sciences, Ko\v{s}ice, Slovak Republic\\
$^{61}$ Institute of Physics, Homi Bhabha National Institute, Bhubaneswar, India\\
$^{62}$ Institute of Physics of the Czech Academy of Sciences, Prague, Czech Republic\\
$^{63}$ Institute of Space Science (ISS), Bucharest, Romania\\
$^{64}$ Institut f\"{u}r Kernphysik, Johann Wolfgang Goethe-Universit\"{a}t Frankfurt, Frankfurt, Germany\\
$^{65}$ Instituto de Ciencias Nucleares, Universidad Nacional Aut\'{o}noma de M\'{e}xico, Mexico City, Mexico\\
$^{66}$ Instituto de F\'{i}sica, Universidade Federal do Rio Grande do Sul (UFRGS), Porto Alegre, Brazil\\
$^{67}$ Instituto de F\'{\i}sica, Universidad Nacional Aut\'{o}noma de M\'{e}xico, Mexico City, Mexico\\
$^{68}$ iThemba LABS, National Research Foundation, Somerset West, South Africa\\
$^{69}$ Jeonbuk National University, Jeonju, Republic of Korea\\
$^{70}$ Korea Institute of Science and Technology Information, Daejeon, Republic of Korea\\
$^{71}$ Laboratoire de Physique Subatomique et de Cosmologie, Universit\'{e} Grenoble-Alpes, CNRS-IN2P3, Grenoble, France\\
$^{72}$ Lawrence Berkeley National Laboratory, Berkeley, California, United States\\
$^{73}$ Lund University Department of Physics, Division of Particle Physics, Lund, Sweden\\
$^{74}$ Marietta Blau Institute, Vienna, Austria\\
$^{75}$ Nagasaki Institute of Applied Science, Nagasaki, Japan\\
$^{76}$ Nara Women{'}s University (NWU), Nara, Japan\\
$^{77}$ National Centre for Nuclear Research, Warsaw, Poland\\
$^{78}$ National Institute of Science Education and Research, Homi Bhabha National Institute, Jatni, India\\
$^{79}$ National Nuclear Research Center, Baku, Azerbaijan\\
$^{80}$ National Research and Innovation Agency - BRIN, Jakarta, Indonesia\\
$^{81}$ Niels Bohr Institute, University of Copenhagen, Copenhagen, Denmark\\
$^{82}$ Nikhef, National institute for subatomic physics, Amsterdam, Netherlands\\
$^{83}$ Nuclear Physics Group, STFC Daresbury Laboratory, Daresbury, United Kingdom\\
$^{84}$ Nuclear Physics Institute of the Czech Academy of Sciences, Husinec-\v{R}e\v{z}, Czech Republic\\
$^{85}$ Oak Ridge National Laboratory, Oak Ridge, Tennessee, United States\\
$^{86}$ Ohio State University, Columbus, Ohio, United States\\
$^{87}$ Physics department, Faculty of science, University of Zagreb, Zagreb, Croatia\\
$^{88}$ Physics Department, Panjab University, Chandigarh, India\\
$^{89}$ Physics Department, University of Jammu, Jammu, India\\
$^{90}$ Physics Program and International Institute for Sustainability with Knotted Chiral Meta Matter (WPI-SKCM$^{2}$), Hiroshima University, Hiroshima, Japan\\
$^{91}$ Physikalisches Institut, Eberhard-Karls-Universit\"{a}t T\"{u}bingen, T\"{u}bingen, Germany\\
$^{92}$ Physikalisches Institut, Ruprecht-Karls-Universit\"{a}t Heidelberg, Heidelberg, Germany\\
$^{93}$ Physik Department, Technische Universit\"{a}t M\"{u}nchen, Munich, Germany\\
$^{94}$ Politecnico di Bari and Sezione INFN, Bari, Italy\\
$^{95}$ Research Division and ExtreMe Matter Institute EMMI, GSI Helmholtzzentrum f\"ur Schwerionenforschung GmbH, Darmstadt, Germany\\
$^{96}$ RIKEN iTHEMS, Wako, Japan\\
$^{97}$ Saga University, Saga, Japan\\
$^{98}$ Saha Institute of Nuclear Physics, Homi Bhabha National Institute, Kolkata, India\\
$^{99}$ School of Physics and Astronomy, University of Birmingham, Birmingham, United Kingdom\\
$^{100}$ Secci\'{o}n F\'{\i}sica, Departamento de Ciencias, Pontificia Universidad Cat\'{o}lica del Per\'{u}, Lima, Peru\\
$^{101}$ SUBATECH, IMT Atlantique, Nantes Universit\'{e}, CNRS-IN2P3, Nantes, France\\
$^{102}$ Sungkyunkwan University, Suwon City, Republic of Korea\\
$^{103}$ Suranaree University of Technology, Nakhon Ratchasima, Thailand\\
$^{104}$ Technical University of Ko\v{s}ice, Ko\v{s}ice, Slovak Republic\\
$^{105}$ The Henryk Niewodniczanski Institute of Nuclear Physics, Polish Academy of Sciences, Cracow, Poland\\
$^{106}$ The University of Texas at Austin, Austin, Texas, United States\\
$^{107}$ Universidad Aut\'{o}noma de Sinaloa, Culiac\'{a}n, Mexico\\
$^{108}$ Universidade de S\~{a}o Paulo (USP), S\~{a}o Paulo, Brazil\\
$^{109}$ Universidade Estadual de Campinas (UNICAMP), Campinas, Brazil\\
$^{110}$ Universidade Federal do ABC, Santo Andre, Brazil\\
$^{111}$ Universitatea Nationala de Stiinta si Tehnologie Politehnica Bucuresti, Bucharest, Romania\\
$^{112}$ University of Cape Town, Cape Town, South Africa\\
$^{113}$ University of Derby, Derby, United Kingdom\\
$^{114}$ University of Houston, Houston, Texas, United States\\
$^{115}$ University of Jyv\"{a}skyl\"{a}, Jyv\"{a}skyl\"{a}, Finland\\
$^{116}$ University of Kansas, Lawrence, Kansas, United States\\
$^{117}$ University of Liverpool, Liverpool, United Kingdom\\
$^{118}$ University of Science and Technology of China, Hefei, China\\
$^{119}$ University of Silesia in Katowice, Katowice, Poland\\
$^{120}$ University of South-Eastern Norway, Kongsberg, Norway\\
$^{121}$ University of Tennessee, Knoxville, Tennessee, United States\\
$^{122}$ University of the Witwatersrand, Johannesburg, South Africa\\
$^{123}$ University of Tokyo, Tokyo, Japan\\
$^{124}$ University of Tsukuba, Tsukuba, Japan\\
$^{125}$ Universit\"{a}t M\"{u}nster, Institut f\"{u}r Kernphysik, M\"{u}nster, Germany\\
$^{126}$ Universit\'{e} Clermont Auvergne, CNRS/IN2P3, LPC, Clermont-Ferrand, France\\
$^{127}$ Universit\'{e} de Lyon, CNRS/IN2P3, Institut de Physique des 2 Infinis de Lyon, Lyon, France\\
$^{128}$ Universit\'{e} de Strasbourg, CNRS, IPHC UMR 7178, F-67000 Strasbourg, France, Strasbourg, France\\
$^{129}$ Universit\'{e} Paris-Saclay, Centre d'Etudes de Saclay (CEA), IRFU, D\'{e}partment de Physique Nucl\'{e}aire (DPhN), Saclay, France\\
$^{130}$ Universit\'{e}  Paris-Saclay, CNRS/IN2P3, IJCLab, Orsay, France\\
$^{131}$ Universit\`{a} degli Studi di Foggia, Foggia, Italy\\
$^{132}$ Universit\`{a} del Piemonte Orientale, Vercelli, Italy\\
$^{133}$ Universit\`{a} di Brescia, Brescia, Italy\\
$^{134}$ Variable Energy Cyclotron Centre, Homi Bhabha National Institute, Kolkata, India\\
$^{135}$ Warsaw University of Technology, Warsaw, Poland\\
$^{136}$ Wayne State University, Detroit, Michigan, United States\\
$^{137}$ Yale University, New Haven, Connecticut, United States\\
$^{138}$ Yildiz Technical University, Istanbul, Turkey\\
$^{139}$ Yonsei University, Seoul, Republic of Korea\\
$^{140}$ Affiliated with an institute formerly covered by a cooperation agreement with CERN\\
$^{141}$ Affiliated with an international laboratory covered by a cooperation agreement with CERN.\\

\end{flushleft}

\end{document}